\documentclass[pra,twocolumn]{revtex4}
\usepackage{graphicx}

\begin{document}
\title{Proposal for optically realizing quantum game}
\author{Lan Zhou and Le-Man Kuang\footnote{Corresponding
author.}\footnote{Email address: lmkuang@hunnu.edu.cn (L. M.
Kuang)}}
\address{Department of Physics, Hunan Normal University, Changsha 410081, People's Republic of China}

\begin{abstract}
We present a proposal for  optically implementing the quantum game
of the two-player quantum prisoner's dilemma involving
nonmaximally entangled states by using beam splitters, phase
shifters, cross-Kerr medium, photon detector and the single-photon
representation of quantum bits.


\noindent PACS number(s): 03.67.-a, 03.67.Lx, 42.50.Dv
\end{abstract}

\maketitle

\section{Introduction}
Recently, much attention has been paid to the topic of quantum
games \cite{meyer,Eisert,Du1,Du2,fli}, which is forming a new area
of study within quantum information and applications of quantum
theory. In additional to their own intrinsic interest, quantum
games open a new window for exploring the fascinating world of
quantum information and quantum mechanics. Actually, various
problems in quantum information and computation, such as quantum
cryptography, quantum cloning, quantum algorithms, can be regarded
as quantum games \cite{simon}. The quantum prisoner's dilemma
\cite{Eisert} is a famous two player quantum game which is
quantization of the so-called Prisoner's Dilemma. Eisert and
coworkers \cite{Eisert} showed that with proper quantum
strategies, the paradox in the classical two-player Prisoner's
Dilemma can be solved under the maximal entangled states, and
discussed  the generalized quantum game of the quantum prisoner's
dilemma where the players share a nonmaximally entangled states.
Then Du and coworkers \cite{Du1} experimentally realized the
quantum game on a nuclear magnetic resonance quantum computer.
Quantum games with multi-players \cite{Du2,simon,ma,iqb} have also
been studied to some extent. It has been found that through
properly choosing quantum strategies, the players may gain more
payoffs in the quantum case than in classical case, and quantum
games can exhibit certain forms of nonclassical equilibria since
quantum entanglement is introduced.

As well known, optical realization is one of the most effective
methods for quantum information processing. The important
experimental implementation of quantum cryptography \cite{Jen},
quantum teleportation \cite{Pan}, quantum dense coding
\cite{Mattle,li} and quantum computation \cite{Knill}, is
completed in quantum optical systems which generally consist of
beam splitters, phase shifters, mirrors, Kerr medium, and so on.
In this paper we add quantum games to the list. The purpose of the
present  paper is to present an optical realization of the quantum
prisoner's dilemma which involves nonmaximally entangled states.
This paper is organized as follows. In  Sec. II, we briefly review
the quantum game of the quantum prisoner's dilemma with
nonmaximally entangled states. In Sec. III and IV, we present
optical preparation of the initial state and optical realization
of the strategic operations in the quantum game. Finally, we
summarize our results in Sec. VI.

\section{Quantum game with nonmaximally entangled states}

Before going into the optical realization, let us briefly review
the quantum game of the two-player quantum prisoner's dilemma with
nonmaximally entangled states \cite{Du1}. There are two players,
and the players have two possible strategies: cooperate
($\hat{C}$) and defect ($\hat{D}$). The payoff table for the
players is shown in Table I. Classically the dominant strategy for
both players is to defect (the Nash equilibrium) since no player
can improve his/her payoff by unilaterally changing his/her own
strategy, even though the Pareto optimal is for both players to
cooperate. This is the dilemma. In the quantum version involving
nonmaximally entangled states, (see Fig. 1), the game is modelled
by two qubits, one for  each player, with the basis states denoted
by  $|C\rangle$ and $|D\rangle$. The physical model of the quantum
game consists of three ingredients: a entangling source of two
qubits, a set of physical instruments that enables the players to
manipulate his or her own qubits in a strategic manner, and a
physical measurement device which determines  the players' pay of
from the outstate of the two qubits. Starting with the product
state $|C C\rangle$ of the two players one acts on the state with
the entangling gate $\hat{J}$ to obtain an entangled state given
by
\begin{equation}
\label{1}
|in\rangle=\cos\left(\frac{\gamma}{2}\right)|C\rangle|C\rangle+i\sin\left(\frac{\gamma}{2}\right)|D\rangle|D\rangle,
\end{equation}
which acts as the initial state of the quantum game. Here $\gamma$
measures the entanglement of the initial state, it changes from
$0$ (no entanglement) to $\pi/2$ (maximal entanglement).

The two players now act with local unitary operators $\hat{U}_A$
and $\hat{U}_B$  on their qubits. Finally, the disentangling gate
$\hat{J}^{\dagger}$ is carried out  and the system is measured in
the computational basis, giving rise to one of the four outcomes
$|C C\rangle$, $|C D\rangle$, $|D C\rangle$, and $|D D\rangle$. If
one allows quantum strategies of the form
\begin{eqnarray}
\label{2}&&\hat{U}(\theta, \phi)=\left(
  \begin{array}{cc}
   e^{i\phi}\cos\frac{\theta}{2} & \sin\frac{\theta}{2} \\
   -\sin\frac{\theta}{2} & e^{-i\phi}\cos\frac{\theta}{2}
  \end{array}
 \right),
\end{eqnarray}
where $0\leq \theta \leq \pi$ and $0\leq \phi \leq \frac{\pi}{2}$,
the two classical strategies $C$ and $D$ is correspond to $U(0,0)$
and $U(\pi,0)$ given by
\begin{equation}
\label{3} \hat{C}=\hat{U}(0,0), \hspace{0.5cm}
\hat{D}=\hat{U}(\pi,0).
\end{equation}

It is worthwhile to mention that Eq. (\ref{2}) is a restriction of
the possible unitary strategies. It is a subset of the possible
strategies. This subset is being chosen because of the ease of
physical implementation.

\begin{table}[htbp]
 \caption{Payoff matrix for the prisoner's dilemma. The first entry in the parenthesis denotes the payoff of Alice and
 the second number is Bob's payoff}
 \begin{ruledtabular}
  \begin{tabular}{|l|cc|}
            & Bob: C & Bob: D \\
   \hline
   Alice: C & (3,3)  & (0,5)  \\
   Alice: D & (5,0)  & (1,1)  \\
  \end{tabular}
 \end{ruledtabular}
\end{table}

Du and coworkers \cite{Du1} showed that the game exhibits an
intriguing structure as a function of the amount of entanglement
with two thresholds $\gamma_1=\arcsin\sqrt{1/5}$ and
$\gamma_2=\arcsin\sqrt{2/5}$ which separate a classical region, an
intermediate region, and a fully quantum region when $\gamma$
changes from $0$ (no entanglement) to $\pi/2$ (maximal
entanglement). $0\le\gamma\le\gamma_1$ is the classical region in
which the quantum game behaves classically, i.e., the only Nash
equilibrium is $\hat{D}\otimes\hat{D}$ and the payoffs for both
players are one, which is the same as in the classical game. The
intermediate region is given by $\gamma_1\le\gamma\le\gamma_2$. In
this region the quantum game can not resolve the dilemma. The
region of $\gamma>\gamma_2$ is the fully quantum region in which a
unique and new Nash equilibrium is only a  Nash equilibriumin the
space of strategies given by Eq. (\ref{2}), which has the property
of being Pareto optimal. In the next two sections, we will give
rise to optical realization of the players' strategic operations
and the $\hat{J}$-gate operation, respectively.

\begin{figure} [htbp]
  \includegraphics[width=8cm]{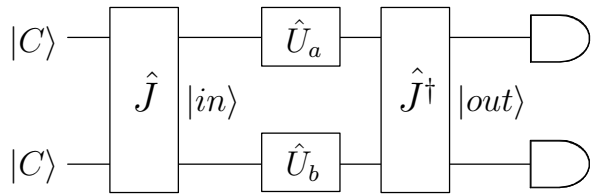}
  \caption{The setup of a two-player quantum game. }
  \label{cong:fig1}
\end{figure}

\section{Realization of quantum strategic operations}
In this section we propose a method to realize the player's
strategy of the quantum prisoner's dilemma with nonmaximally
entangled state only by 50:50 lossless beam splitters, phase
shifters and mirrors, which are easy to operate for players.

A typical two-mode mixer that preserves the total number of
photons in the mode pair, is a lossless beam splitter with
transparency $T$. Denoting the input annihilation operators as
$\hat{a}$ and $\hat{b}$, and the output's as $\hat{a}'$ and
$\hat{b}'$. The beam splitter transform the input $\hat{a}$ and
$\hat{b}$ into the output $\hat{a}'$ and $\hat{b}'$ as following
\begin{eqnarray}
\label{4}\left(
  \begin{array}{c}
   \hat{a}' \\
   \hat{b}'
  \end{array}
 \right)=\left(
  \begin{array}{cc}
   \cos\frac{\theta}{2} & -i\sin\frac{\theta}{2}\\
   -i\sin\frac{\theta}{2} & \cos\frac{\theta}{2}
  \end{array}
 \right)\left(\begin{array}{c}
   \hat{a }\\
   \hat{b}
  \end{array}\right),
\end{eqnarray}
where the transparency $T=cos^{2}(\theta/2)$. this is one ways to
relate the input state to the output state. A beam splitter also
can be described by a unitary operator with the form
\begin{equation}
\label{5}
\hat{B}(\theta)=\exp{\left[-i\theta(\hat{a}^{\dagger}\hat{b}+\hat{b}^{\dagger}\hat{a})\right]}.
\end{equation}

\begin{figure} [htbp]
  \includegraphics[width=6cm]{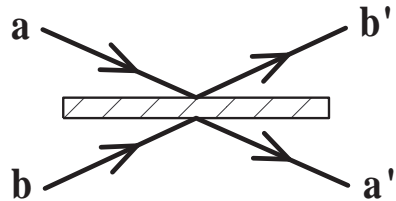}
  \caption{A beam splitter with two inputs $a,b$, and two outputs $a',b'$.}
  \label{cong:bs}
\end{figure}

If impinging on a beam splitter with one photon in one mode and
zero photon in the other, the output lights are entangled through
the following expressions
\begin{eqnarray}
\label{6}
\hat{B}(\theta)|1\rangle_{a}|0\rangle_{b}=\cos\frac{\theta}{2}|1\rangle_{a}|0\rangle_{b}
-i\sin\frac{\theta}{2}|0\rangle_{a}|1\rangle_{b},
\nonumber \\
\hat{B}(\theta)|0\rangle_{a}|1\rangle_{b}=\cos\frac{\theta}{2}|0\rangle_{a}|1\rangle_{b}
-i\sin\frac{\theta}{2}|1\rangle_{a}|0\rangle_{b}.
\end{eqnarray}

In the above single-photon representation of quantum bits, let
$|C\rangle=|1\rangle|0\rangle$, $|D\rangle=|0\rangle|1\rangle$,
under the beam-splitter transformation we have
\begin{eqnarray}
\label{7}
\hat{B}(\theta)|C\rangle=\cos\frac{\theta}{2}|C\rangle-i\sin\frac{\theta}{2}|D\rangle, \nonumber \\
\hat{B}(\theta)|D\rangle=\cos\frac{\theta}{2}|D\rangle-i\sin\frac{\theta}{2}|C\rangle,
\end{eqnarray}
where if we set $\theta=\pi/2$, it is a 50:50 beam splitter with
transparency $T=1/2$.

The second device we should introduce is a phase shifter in mode $a$, which can be described by the unitary operate
\begin{equation}
\label{8}
\hat{P}(\phi)=\exp{\left[i\phi\hat{a}^{\dagger}\hat{a}\right]}.
\end{equation}

Putting a phase shifter in mode $a$ and mode $b$, respectively,
with the phase shifter in mode $b$ conjugate to the one in mode
$a$, for convenience, we denote
$\hat{P}(\phi,-\phi)=\hat{P}_{a}(\phi)\hat{P}_{b}(-\phi)$
\begin{eqnarray}
\label{9}
\hat{P}(\phi,-\phi)|C\rangle&=&e^{i\phi}|C\rangle, \nonumber \\
\hat{P}(\phi,-\phi)|D\rangle&=&e^{-i\phi}|D\rangle,
\end{eqnarray}
which correspond to the rotation around the $z$ axes by $\phi$.
The rotations around the $y$ and $x$ axes by $\theta$ and $\eta$,
respectively, can be realized by the following form
\begin{eqnarray}
\label{10}
\hat{U}_{y}\left(\frac{\theta}{2}\right)=\hat{B}\left(-\frac{\pi}{2}\right)\hat{P}\left(-\frac{\theta}{2},\frac{\theta}{2}\right)\hat{B}\left(\frac{\pi}{2}\right),  \\
\label{11}
\hat{U}_{x}\left(\frac{\eta}{2}\right)=\hat{P}\left(-\frac{\pi}{4},\frac{\pi}{4}\right)\hat{U}_{y}\left(-\frac{\eta}{2}\right)
\hat{P}\left(\frac{\pi}{4},\frac{\pi}{4}\right),
\end{eqnarray}
where the rotations around the $y$ and $x$ axes are defined by
\cite{campos}
\begin{eqnarray}
\label{11'}
\hat{U}_{x}(\alpha)&=&\exp\left[-\frac{i\alpha}{2}(\hat{a}^{\dagger}\hat{b}
            + \hat{b}^{\dagger}\hat{a})\right], \\
 \hat{U}_{y}(\beta)&=&\exp\left[-\frac{\beta}{2}(\hat{a}^{\dagger}\hat{b}
            - \hat{b}^{\dagger}\hat{a})\right].
\end{eqnarray}

\begin{figure} [htbp]
  \includegraphics[width=7cm]{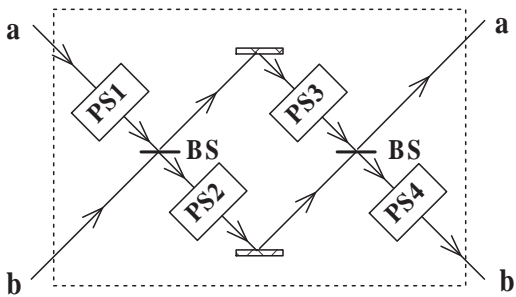}
  \caption{Schematic setup for implementing the quantum strategy $\hat{U}(\theta,\phi)$. It consists of two 50:50 beam splitters, two mirrors, and four
  phase shifters. Among these phase shifters, PS2 acts conjugate to PS3 and PS1 acts conjugate to PS4. The two beam
   splitters and two mirrors along with PS2 and PS3 realize the $\hat{U}_{y}(-\theta/2)$.}
  \label{cong:strategy}
\end{figure}

Then the unitary operator of the player's strategy shown in
Eq.(\ref{1}), can be realized using the following identity:
\begin{eqnarray}
\label{12}\hat{U}(\theta,
\phi)=\hat{P}(0,-\phi)\hat{U}_{y}\left(-\frac{\theta}{2}\right)\hat{P}(\phi,0),
\end{eqnarray}
and the experiment setup is the combination of the beam splitters,
phase shifters and mirrors shown in Fig.\ref{cong:strategy}.

\section{Realization of the $\hat{J}$-gate operation}

The $\hat{J}$ gate is the most important operation in the quantum
game since it introduces quantum entanglement. In this section we
show that the  $\hat{J}$ gate can be realized in terms of a
sequence of beam splitters, cross Kerr medium, and phase shifters.
Unitary transformations corresponding to beam splitters and phase
shifters are given by Eqs. (\ref{3}) and (\ref{6}), respectively,
while a unitary transformation which characterizes a cross-Kerr
medium is given by
\begin{equation}
\label{13'} \hat{K}(\chi)=\exp{\left[-i\chi
\hat{a}^{\dagger}\hat{a}\hat{b}^{\dagger}\hat{b}\right]},
\end{equation}
where the nonlinear Kerr coefficient $\chi$ is proportional to the
third order susceptibility $\chi^{(3)}$ of the medium and the
interaction time within the medium. Here we assume the
self-modulation terms $\hat{a}^{\dagger 2}\hat{a}^{2}$ and
 $\hat{b}^{\dagger 2}\hat{b}^{2}$ can be ignored by an appropriate
 choice of resonance \cite{Gerry1,imoto,Gerry2}.

\begin{figure} [htbp]
  \includegraphics[width=6cm]{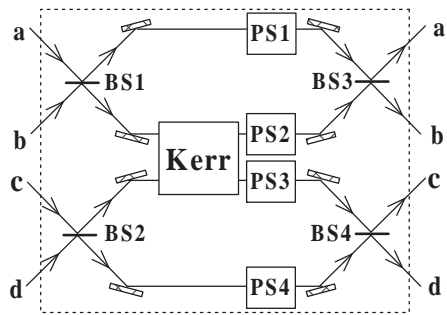}
  \caption{Schematic setup for implementing the $\hat{J}$-gate. The cross-Kerr medium couples mode $a$ and $d$. All the beam
  splitters here are 50:50 lossless beam splitters, phase shifter PS1 is conjugate to PS2 with parameter
  $\theta_{1}$, and PS3 is conjugate to PS4 with parameter $\theta_{2}$, whose values is connected with the nonlinear
  Kerr coefficient.}
  \label{cong:jgate}
\end{figure}

Our scheme to realize the $\hat{J}$-gate is indicated in
Fig.\ref{cong:jgate}, which consists of four 50:50 beam splitters,
four phase shifters, one cross-Kerr medium and some mirrors. All
of the BSs are 50:50 beam splitters. The cross-Kerr medium couples
the lights in mode $a$ and $d$. The lights of Four modes first
pass through BS1 and BS2, and two of them go through the
cross-Kerr medium, then they change their phases by four phase
shifters, whose value is connected with the nonlinear Kerr
coefficient, at last they propagate through beam splitters, BS3
and BS4.

Then the unitary transformation to realize the $\hat{J}$-gate can
be expressed as the product of a sequence of beam-splitter
transformations, the cross-Kerr transformation and phase-shifter
transformations, which is given by
\begin{eqnarray}
\label{13}
\hat{J}&=&\hat{B}_{4}\left(-\frac{\pi}{2}\right)\hat{B}_{3}\left(-\frac{\pi}{2}\right)
\hat{P}(\theta_{2},-\theta_{2})\hat{P}(\theta_{1},-\theta_{1})
\nonumber\\
& & \times \hat{K}(\gamma)
\hat{B}_{2}\left(\frac{\pi}{2}\right)\hat{B}_{1}\left(\frac{\pi}{2}\right),
\end{eqnarray}
which can be further reduced to the following simple form
\begin{equation}
\label{14}
\hat{J}=\exp{\left[-i\frac{\gamma}{4}\left(\hat{a}^{\dagger}\hat{b}-\hat{a}\hat{b}^{\dagger}\right)
\left(\hat{c}^{\dagger}\hat{d}-\hat{c}\hat{d}^{\dagger}\right)\right]}.
\end{equation}

If we assume that the input state of the setup indicated in
Fig.\ref{cong:jgate} is  $|C C\rangle$, then its out state is
given by
\begin{equation}
\label{15}
|\psi_{0}\rangle=\exp{\left[-i\frac{\gamma}{4}\left(\hat{a}^{\dagger}\hat{b}-\hat{a}\hat{b}^{\dagger}\right)
\left(\hat{c}^{\dagger}\hat{d}-\hat{c}\hat{d}^{\dagger}\right)\right]}|C
C\rangle,
\end{equation}
which can be explicitly written as
\begin{equation}
\label{16}
|\psi_{0}\rangle=\cos\frac{\gamma}{2}|C\rangle|C\rangle+i\sin\frac{\gamma}{2}|D\rangle|D\rangle.
\end{equation}
This is just the entangled state as the initial state of the
quantum game $|in \rangle$ given by (\ref{1}).

It is easy to verify the following commutators
\begin{eqnarray}
\label{17} &[\hat{J}, \hat{D}\otimes\hat{D}]&=0, \hspace{0.5cm}
[\hat{J}, \hat{D}\otimes\hat{D}]=0,\nonumber \\
&[\hat{J}, \hat{C}\otimes\hat{D}]&=0,
\end{eqnarray}
which are the subsidiary conditions which the entangling operation
$\hat{J}$ must be satisfied in order to guarantee that the
ordinary prisoner's dilemma is faithfully represented. Hence,
 the unitary transformation (\ref{14}) faithfully realize the
 $\hat{J}$-gate operation to generate the initial entangled state
 (\ref{1}).

\section{Conclusion}
In this paper we have presented a optical scheme to realize the
quantum prisoner's dilemma by using beam splitters, phase
shifters, cross-Kerr medium and mirrors with single-photon
sources. It should also be possible to use this method to realize
multi-player quantum game. We believe that the practical
realization of such quantum games should provide further insight
into the studies of quantum networks.

\acknowledgments
This work was supported in part the National
Fundamental Research Program (2001CB309310), the National Natural
Science Foundation of China under Grant Nos. 90203018 and
10075018, the State Education Ministry of China and the
Educational Committee of Hunan Province.

\end{document}